# CLOUD NETWORK MANAGEMENT MODEL - A NOVEL APPROACH TO MANAGE CLOUD TRAFFIC

Dr. Mamta Madan[1] and Mohit Mathur[2]

1Professor
Department of Computer Science
Vivekananda Institute of Professional Studies,
(Affiliated to Guru Gobind Singh Indraprastha University), Delhi, India

2Assisstant Professor
Department of Information Technology
Jagan Institute of Management Studies,
(Affiliated to Guru Gobind Singh Indraprastha University), Delhi, India

## ABSTRACT

*Cloud is in the air. More and More companies and personals are connecting to cloud with so many variety of offering provided by the companies. The cloud services are based on Internet i.e. TCP/IP. The paper discusses limitations of one of the main existing network management protocol i.e. Simple Network Management Protocol (SNMP) with respect to the current network conditions. The network traffic is growing at a high speed. When we talk about the networked environment of cloud, the monitoring tool should be capable of handling the traffic tribulations efficiently and represent a correct scenario of the network condition. The proposed Model 'Cloud Network Management Model (CNMM)' provides a comprehensive solution to manage the growing traffic in cloud and trying to improve communication of manager and agents as in SNMP (the traditional TCP/IP network management protocol). Firstly CNMM concentrates on reduction of packet exchange between manager and agent. Secondly it eliminates the counter problems exist in SNMP by having periodic updates from agent without querying by the manager. For better management we are including managers using virtualized technology. CNMM is a proposed model with efficient communication, secure packet delivery and reduced traffic. Though the proposed model supposed to manage the cloud traffic in a better and efficient way, the model is still a theoretical study, its implementation and results are yet to discover. The model however is the first step towards development of supported algorithms and protocol. Our further study will concentrate on development of supported algorithms.*

## 1. INTRODUCTION

The Internet is growing by providing lots of online services like search engines, banking, social networking, gaming and video conferencing across multiple locations. In recent years, large investments have been made in massive data centers supporting computing services, by Cloud Service Providers (CSPs) such as Face book, Google, Microsoft, and Yahoo! [8]. The significant investment in capital outlay by these companies represents an ongoing trend of moving applications, e.g., for desktops or resource-constrained devices like smart phones, into the cloud.

Cloud is on the hype. It is expanding day by day, but the extent of this growth has not been discussed from a technical point. The cloud traffic growth is a consequence of the fast adoption and migration to cloud .Moreover the migration to cloud is due to the ability of cloud data centers to handle higher traffic loads. These data centers use virtualization and automation. Thus data centers have increased performance, higher capacity and great throughput. According to reports from Cisco Systems, Global cloud (Compound Annual Growth Rate, CAGR) traffic is expected to grow 4.5-fold – a 35% combined annual growth rate. The traffic between the data centers and beyond data centers is increasing rapidly. [1] Figure 1 shows the growing traffic statistics in data centers till 2017. [1]

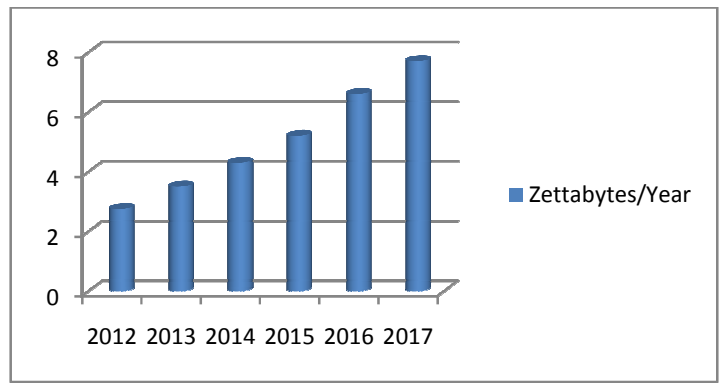

Figure 1 Data Center Traffic Growths

Networking capabilities plays a crucial role for getting data from and storing data to the cloud. The networking capabilities include networking devices, bandwidth, protocols etc.  Internet carries several types of traffic. Growing Internet and Cloud are major contributors to this traffic. The Internet community and researchers are making their best effort to reduce or optimize the traffic conditions. Though networking companies are tiring with development of extremely high speed networking devices, there still needs some methods to reduce or efficiently manage the traffic on the cloud. The problem lies with the protocols. The traditional TCP/IP protocols are not being able to cop up with the level of services that cloud requires. When a packet travels on the cloud it may pass through several components before it ever leaves the system: system buffer, transfer application, network stack, software VPN, software firewalls and filters, network drivers, and the hardware network adapter, Network Devices such as Routers, and Gateways. Each device adds its jitter in processing the packet. Hence there is very less scope in making delivery fast. The Internet Community has already developed so many protocols like Multi Protocol Label Switching (MPLS), Resource Reservation Protocol (RSVP) etc that will help in timely delivery of packets with required quality of services. In our study we are concentrated on how to reduce and bitterly manage the network traffic. For that we surveyed the basic Network management protocol Simple Network Management Protocol (SNMP) which is widely adopted. During the study, we identified SNMP as one of the protocols whose communication can be optimized for faster and better cloud network; moreover the traffic generated by SNMP can be reduced to some extent.

For better Quality of Services (QOS), system/network administrators should be always aware of the current status of the networking devices called agents, including their CPU loads, storage usage etc. Currently, SNMP has been widely used in remote monitoring of network devices and hosts. In this paper, we would like to discuss the weakness of SNMP in management communication as well as we will introduce a new Model that will try to overcome those problems. Management usually requires the support of an agent in the managed host, and the database in the agent provides the management information needed for a management application. Let us first enumerate the problems that lie with traditional SNMP protocol.

## 2. PROBLEMS WITH SNMP

In SNMP we know that the Network Management Station (NMS) called manager periodically requests or polls the agent. The MIB inside agents contains a counter that counts number of bytes transmitted and received in a particular time interval on each of its interfaces. The counter is cyclic. The SNMP counters counts only a running total and not the count the number of packets per interval. SNMP manager send polls to agent to compute packets per interval in short duration of time. SNMP polls after every five minutes. Thus SNMP poller periodically records these counters and collects information.[6]

SNMP data collected by polling has many known limitations.

- SNMP uses unreliable User Datagram Protocol (UDP) transport, Data may be lost in transit .
- Sometimes an SNMP poller restarts and it loses its track of a counter, counter resets (say after a router reboot), which results in large error in the estimate of traffic. In early versions of SNMP 32-bit counters were used and these counters reset quickly on high speed links. Sometimes SNMP poller wrongly calculates the average rate as per information received, ignoring the missing polls. [12]
- "Jitter" caused by polling is another problem in SNMP. The Network Management Station must perform polls to many devices and these polls cannot be performed concurrently. These query –reply packets take some time to transit the network [9].Finally the result is that the reply packets reach late due to this jitter. Moreover routers give low priority to SNMP packets; therefore they have a delayed response.
- SNMP processes on agents  are given low-priority and hence they  have a delayed response;
- SNMP is too periodic. Sometimes polling cycles from 30 seconds to several minutes long does not produce the actual picture of the network routing conditions. Even if we speed up the polling cycle it would miss many routing state changes, and would generate  much management traffic overhead[12].
- SNMP communication delays the action to be taken by manager, as manager has to first send a query message in which it has to access the MIB , the object data then travel all the way to manager and if required send the update message to manager. Thus using SNMP is not meant for very large networks because sending a packet to get another packet causes delay in communication and hence in management. This type of polling causes large volumes of regular messages and end in problem response times that may be unacceptable [7].
- There is no acknowledgement for Trap messages in SNMP. If UDP is used with Internet Protocol (IP) to deliver trap message by agent, the agent gets no response whether the trap message has been delivered to manager or not. This is unacceptable for such critical messages.
- SNMP does not directly support crucial commands. The only way to prompt an event at an agent is indirectly by setting an object value. A more efficient way is to use remote procedure calls with parameters, conditions, status and results, that SNMP does not support.

- SNMP marginal errors should not be ignored as feeding such small errors into management process causes major problems, corrupting the results and leads to poor management.

## 3. Cloud Network Management Model (CNMM)-A Novel Approach to Manage Cloud Traffic

The Model is based on the agent manager relationship.

### 3.1 Entities involved in the CNMM

- Virtualized Network Management Server (Manager): is used to manage and supervise the entire network. It receives all the information and displays it. It may be a pool of virtualized servers. We can take cloud services for Obtaining Manager Services. We assume that Manager is virtualized pool of servers kept on cloud. The Manager is usually in listening mode to have updates from the agents [2].
- The Network Management Agent (Agent): A network node that contains a CNMM agent. These agents collect and store management information. The agent then creates the required information send update to Manger. Managed devices (Agents), sometimes called network elements, are mostly routers having special software installed in them. They keep the information in the database having collective information from there routing tables regularly updated through routing protocols and regularly send the updates to the manager [3]. While implementing CNMM two important points need to be considered regarding agents. Firstly, we know that the agents are usually routers and routers are busy with high traffic. The implementations of CNMM will further affect the performance of router i.e the implementation of CNMM requires generating update packets and sending updates at regular interval which in turn make router processor busy. The solution to this problem is that the model suggests sending updates at regular but large intervals and if the situation is unmanageable within that interval the agent forwards a trap message. The large interval here depends on the implementation of model. Moreover we should remember that the model saves the processing overhead required for query packets that SNMP generates. Secondly, as we know that the management traffic given lower priority over user data traffic such as voice, chat etc, to solve the problem the model provide options to prioritize CNMM traffic The prioritization of CNMM traffic will be discussed in our future work.
- Management information base (MIBs). The agent keeps information in Management information base (MIB). This information is a collection of objects or data values. Here each agent will keep the management information about number of packets sent / received. The SNMP agent process prepares this information from the raw data collected by it about number of bytes sent/ received [14]. It sends the update packets by extracting information from MIB. Hence here we are eliminating the problem of actual data required by manager i.e. number of packets sent/ received. Moreover our MIB is motivated from the routing table kept by agent. The agent uses this dynamic routing information from routing table to prepare network condition summary and packet information in form of packets sent and received from bytes sent/ received. This will represent the real picture of the network condition to the manager.

### 3.2 Working of CNMM

The overall working involves a set of agents sending updates about their performance to managers in the cloud. The updates will be sending only when the value of any parameter of an agent goes below its threshold level or when timer expires. However a manager may also send a query packet if it does not listen from agent for a long time .Whenever an agent sends an update packet to manager any of the virtualized manager machines reply by checking all parameters. The benefit of virtualized manager machine is that we are making an efficient use of manager machines. Moreover since a large number of agents will send their update in short period of time, it will be difficult to handle them by a single (Non Virtualized) machine. Now if we look at the information contained in update packet will be the number of packets sent/received instead of number of bytes sent/ received in a particular interval of time. Each agent keeps its performance or other information in a set of objects called Management Information Base (MIB). The manager has rights to access/modify these MIB's. Though each time to access the Agents MIB manager has to authenticate and show access rights to the agent. Usually agent will take initiative by sending the update message to manager but manager may need to access MIB while responding to these updates to modify the object values. The initiative taken by agent to update manager will reduce the unnecessary traffic created by SNMP in Query and reply packets. This will also eliminate polling problems that lies in SNMP. Moreover it will reduce the time of overall communication and further eliminate the problems related to counters and jitter.

For better QoS of cloud services, system/network administrators of network should also be always aware of the current status of the manager machines in the cluster, including their CPU loads, storage usage, and network utilization. Furthermore, administrators are also interested in how many Virtual Machine (VM) instances are allocated in a manager host machine and how well each VM instance is running. As a great number of managers that are deployed to provide virtual machines to a variety of agents. The Model is a hybrid of centralized and decentralized management. The basis of suggested Model lies in the initiative taken during the communication. The initiative to transfer is taken by agent instead of manager. The manager will not generate any request messages as in SNMP. The agent keeps database of information such as MIB and send updates of its database to its manager. Some information might relate to the system, some might be network related, some might be resource specific and there will be events associated with each incident [10]. The updates are sent similarly as in link state protocol of routing i.e. the agent send an update whenever the value of any object goes below its threshold. Each object in the database has been defined with a threshold value. The value below that threshold will not be tolerated and immediately informed to the manager to take action. The communication will be initiated by agent instead of manager. For proper management, all the messages in CNMM are acknowledged.

### 3.3 Types of Messages

CNMM defines the following basic types of packets:

- Regular Update Packet( From Agent To Manager)

- Trap Message(from Agent to Manager, in case of urgent action)

- Action/ Set Message(Reply Message From manager to Agent)

- Get Message(From Manager to agent in case manager does not hear from agent for a long time)

- Advertisement Message (By new Agent that enters a network to Manager.)

- Registration Message(By Manager to agent after receiving advertisement)

International Journal on Cloud Computing: Services and Architecture (IJCCSA) ,Vol. 4, No. 5, October 2014

### 3.3.1  Regular Update Packet & Action/ Set Message

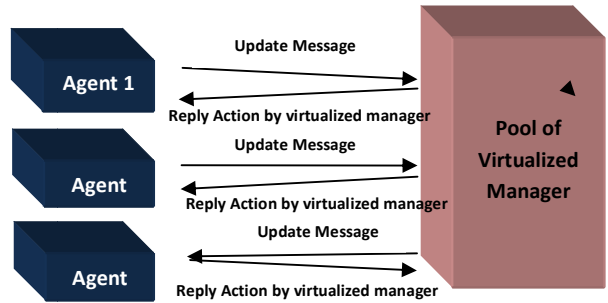

Figure 2 A simple example of CNMM communication

The Figure 2 shows how update message is communicated between agents and pool of virtualized manager. Agent keeps its performance value (Packet sent/ received and other related information) as set of object in a database; the values will be updated regularly as per performance of the agent. The values will be monitored by the program inside the agent itself and compared with the minimum threshold value below which it cannot be tolerated. Moreover a value called minimum value which is above threshold level is also defined. These values are defined for various performance parameters of agent. This means that an agent will keep multiple values of its performance in its database. The messages are generated by agent whenever the performance value of any parameter of agent reaches minimum level or before the expiration of update timer described later. Such messages are then forwarded to the respective manager to take some action. This means that manager can assume that everything is going well if it does not get any packet from a manager up to a period of time. For this manager keeps a timer for each agent. If the timer expires the manager sends get message the agent to check whether everything is fine.

More over it may happen that the value of the agent performance may reach below threshold level. In this case the agent generates an alert message. The manager then takes care of such messages by read or writes instructions/ messages sent to agent. Figure 3 shows the levels at which update and alert messages were sent to the manager by agent. Manager has all rights to update or edit or access the values inside agent database but it has to authenticate itself before messages are accepted by to agent database. Our Model reduces the management traffic by avoiding the query packets generated by SNMP managers.

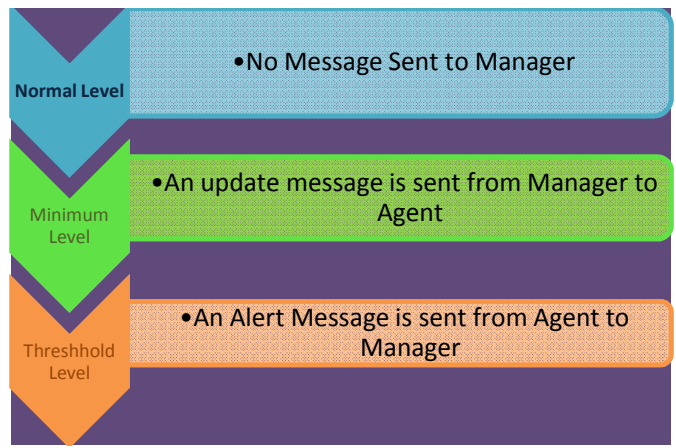

Figure 3 Agents Performance levels in CNMM

### 3.3.2 Trap & Reply Message

An Agent can also send a trap message in case of emergency i.e. when it requires an immediate action from manager without delay. The manager in that case will process the request with highest priority. Figure4 shows exchange of trap and trap reply message. The messages are generated by agent whenever the performance value of any parameter of agent reaches threshold message.

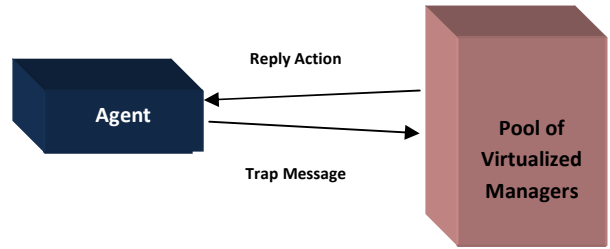

Figure 4 Trap & Reply Message

### 3.3.3 Advertisement and Registration

CNMM also support advertisement and the registration. Whenever a new host enters the network it first of all discovers its NMS and then send an advertisement message to the NMS about its existence in the network. The NMS then registers the host. The extension of this capability is to extend the same to a broader cloud scenario without compromising the functioning and overhead. Figure 5 shows the exchange of advertisement and registration message.

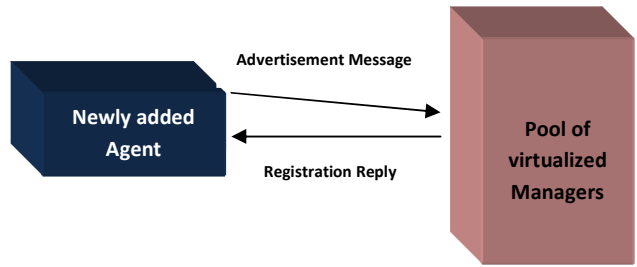

Figure 5 Advertisement & Registration Message

### 3.4 CNMM Timers

**Update Timer**

For a proper management, Manager keeps timers. Manager keeps an update timer which starts when manager receives an update message from agent. The agent is expected to send an update before expiration of this timer. When the timer expires the manager sends a get message to get the status of agent. If the agent replies, all goes well otherwise another get message is sent. This is repeated three times after which the manager alerts the management console about the event to take some action.

### 3.5 Virtualized Manager

Next the Model suggests that we may have a pool of managers that work for a set of agents. But these managers are using virtualization technology. A physical manager is converted into

1515

multiple virtual machines using virtualization technology. Each virtual manager acts like a unique physical device, capable of running its own operating system (OS). [13] Our future work will focus on how this virtualization technology works and will also highlight manager to manager communication, packet formats etc. Making use of Virtualized Manager carries several benefits like increase utilization of infrastructure, fast replying time to agents, application downtime and recovery time reduction, fast deployment of applications and reducing infrastructure and operating costs.[13] Virtualization is realized by introducing a virtualization layer between hardware and the operating system. The virtualization layer, realized by a VM hypervisor or Virtual Machine Monitor (VMM), enables the creation of virtual machines in different operating systems, and allows their shared access to the real hardware resources, including CPU, memory, and I/O devices. Further we can insert a thin layer of software (hypervisor) between the server hardware and the operating system. The hypervisor contains virtual hardware containers that host applications and operating systems [13]. Figure 6 shows the various layers of communication inside virtualized managers. [13]

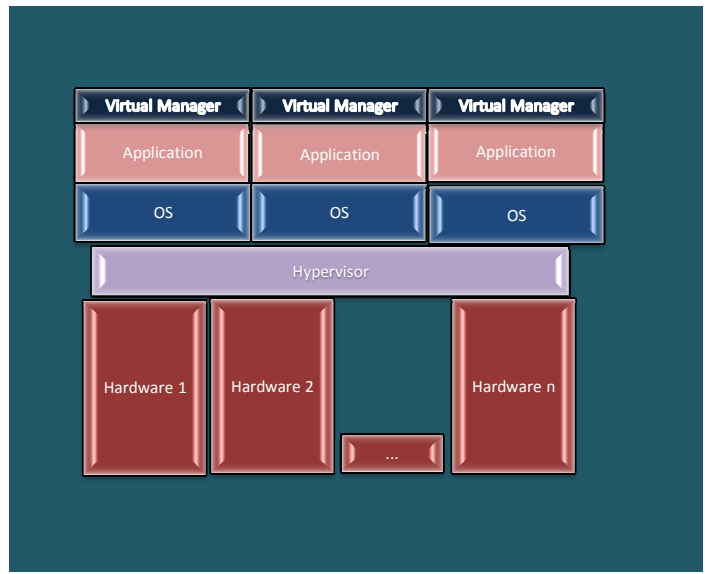

Figure 6 A pools of Virtualized Managers in the Cloud

Use of hypervisor imposes the many benefits on virtual managers that traditional managers do not have, like Isolation, Multiplicity, Abstraction and Encapsulation. [13]

### 3.6 CNMM Security

We know that SNMP v3 uses User security model (USM). The main reason of not deploying USM is that USM utilizes a separate user and key management infrastructure. Deploying another user and key management infrastructure introduces significant operational costs [4]. The SNMP architecture was not designed with session based security in mind. As a consequence, the original ASIs between the subsystems do not pass all the necessary security information to all subsystems.

If we look at the other security solutions for SNMP that leverage existing secure transport protocols such as Secure Shell(SSH), Transport Layer Security(TLS) and Datagram Transport Layer security(DTLS). These protocols have an already widely deployed security infrastructure and key management for these protocols is generally well understood. These all secure transport models have in common that they use a concept of a session and provide security services based on a per session basis, called session-based security. By providing security services at the transport layer instead of embedding security services into the SNMP protocol itself, the usability in operational environments can be significantly improved.

However using SSH/TLS/DTLS require session establishment, however we are more concerned about the speed of communication as well as achieving basic security [15]. Session establishment and other sophistication required by these protocols causes delays, hence we concentrate only on providing authentication, confidentiality and integrity and access control to the CNMM packets. We do not suggest any establishment of session as it will cause delay to overall communication. We used a hybrid approach to achieve complete security. The CNMM security includes Securing (providing confidentiality and Authentication) to the packets that are being exchanged between Agent and manager.

The CNMM secure packet exchange involves authentication, confidentiality and message integrity.

The purpose of the CNMM secure packet exchange involves taking an application message to be transmitted, fragmenting it, encapsulating it with appropriate headers, and finally encrypting it before it is forwarded using UDP protocol.

The steps involved in creating a secure packet are as follows:

**Step 1** First of all a header is added to the application data portion. A header keeps information such as data size and the MAC. The data is then classified into packets.

**Step 2** Packets are then compressed, so that it will be reduced to contain less byte.

**Step 3** The data is then encrypted using encryption techniques. To provide authentication code called message authentication Code (MAC) is calculated and placed in the header. A secret key is created during creation of MAC, this key is either a client chosen MAC secret or a server chosen MAC secret respectively, and it depends on which party prepares the packet.

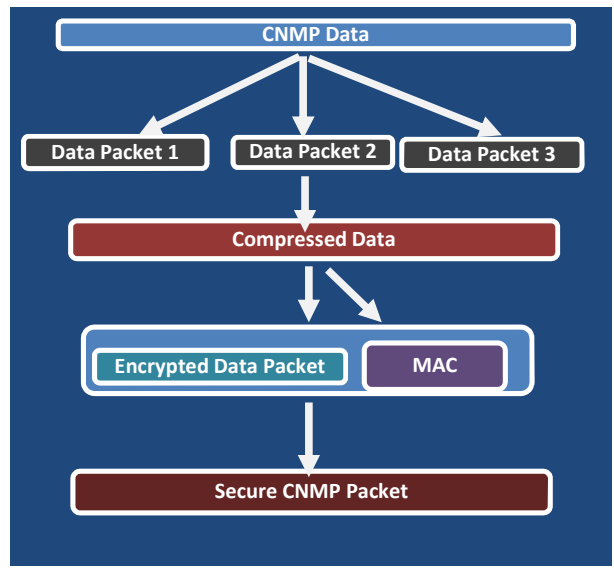

Figure 7 CNMM Application Data Processing for security

**Step 4** Next, the data plus the MAC are encrypted using a previously agreed upon symmetric encryption algorithm, for example Data Encryption Standard (DES), triple DES, International data Encryption Algorithm (IDEA), Blowfish etc. Both data and MAC are encrypted.

**Step 5** Encrypted Data packet and MAC together makes a secure packet as it is ready to move to the insecure public network.

The whole procedure provides confidentiality, authentication, Integrity and compression to the CNMM communication [5].The whole procedure is illustrated in Figure 7.

### 3.7 Benefits of the Proposed Model

1. Help enhance network performance and lower risk- Since packets are initiated by CNMM Agent, The polling done in SNMP can be avoided. Less number of packets led to less jitter and hence enhances network performance.
2. Reduce network Traffic- The polling packets in SNMP are not used in CNMM as well as agent too generates a packet only when it is required, this reduces the unnecessary query and response messages generated in SNMP.
3. Secure Communication between Managers and Agents- CNMM packets use secure and authenticated packets which provides confidentiality, authentication and Message Integrity to each packet communicated.
4. Better Communication and updated information with managers- Managers are still updated, though we have less packet exchange. The updates sent by agent are kept in manager's database.
5. Faster recovery in case of failures- since we are using the concept of manager virtualization the recovery to failures is very fast.
6. Virtualization benefits- All benefits that come under virtualization technology were achieved in CNMM Manager.
7. No Polling Problems- All the problems related to Polling in SNMP will be solved as no polling is done by manager to gather information / status of CNMM Agent.
8. More Accurate Change Management and Planning Processes: Since CNMM's agent MIB is motivated with the Routing Table information; it calculates and the change in routing and traffic based on the traffic and routing matrix which is more close to the actual network scenario. The new traffic and routing picture and its analysis show whether any congestion will result or not. [11]

## 4. CONCLUSION

Cloud computing is an emerging technology. More and More individuals and companies are adopting cloud at a faster rate, due to which internet traffic is increasing at a pace which is difficult to manage. With development of new technologies in the cloud we need to modify the traditional protocols to manage the increasing cloud traffic. Cloud Network Management Model is such a Model that efficiently manages cloud traffic with more accurate results and providing security to the management packets being exchanged.  Though, the paper describes the Cloud Network Management Model at abstract level. Our future research work will concentrate on describing each part of the Model at detailed design and at implementation level. In continuation we will focus on analyzing the use of OpenFlow to check the flows of traffic in cloud and then take some decisions on to which network manager to forward the messages to.

## ACKNOWLEDGMENT

First of all we would like to acknowledge Goddess Saraswati for making us capable of writing this research paper. Further, we would like to thank everyone at workplace and anonymous reviewers for their useful comments and suggestions.

## REFERENCES

...

## Authors

**Dr Mamta Madan**

Professor Mamta Madan is an accomplished professor of Computer Science at VIPS, IP University. She has over 17 years of experience in research and academics. She is actively involved in research in the areas of artificial intelligence, software engineering, data mining and cloud computing. She is guiding many Ph.D students enrolled at various Indian universities. She is associated with many professional and research bodies like Central Board of Secondary Education, Computer Society of India etc. Her expertise goes well beyond the classroom, as she is in the panel of examiners at various universities and has evaluated numerous projects of computer science. She has published and presented many papers in National and International Journals of repute.

**Mr. Mohit Mathur**

Mohit Mathur is working as Head of Department, Dept. of Information Technology at Jagan Institute of Management Studies, Delhi, India. He has done his Graduation of Delhi University and MCA from Department of Electronics, Ministry of Information Technology, India. His area of Interest is Network / Network security. He is pursuing his Research work on Cloud Computing. He has already written many research papers in the same area representing different aspects of cloud like security, traffic, scalability, migration etc.